\documentclass[12pt]{article}
\usepackage[english]{babel}
\usepackage[T1]{fontenc}
\usepackage{amsmath,amsfonts,graphicx,color}
\newcommand{\no}{\nonumber }
\newcommand{\be}{\begin{equation}~}
\newcommand{\eeq}{\end{equation}~}
\newcommand{\bee}{\begin{eqnarray}~}
\newcommand{\ee}{\end{eqnarray}}
\newcommand{\been}{\begin{eqnarray*}~}
\newcommand{\een}{\end{eqnarray*}~}
\newcommand{\pa}{\partial}

\begin{document}
\begin{titlepage}
\begin{flushright}
TIT/HEP-659\\
July, 2017
\end{flushright}
\vspace{0.5cm}
\begin{center}
{\Large \bf
ODE/IM correspondence and the Argyres-Douglas theory
}
\lineskip .75em
\vskip 2.5cm
{\large  Katsushi Ito and Hongfei Shu }
\vskip 2.5em
 {\normalsize\it Department of Physics,\\
Tokyo Institute of Technology\\
Tokyo, 152-8551, Japan} 
\vskip 3.0em
\end{center}

\begin{abstract}
We study the  quantum spectral curve of the Argyres-Douglas theories in the Nekrasov-Sahashvili limit of the Omega-background.
Using the ODE/IM correspondence we investigate the quantum integrable model corresponding to the quantum  spectral curve.
We show that the  models for the $A_{2N}$-type theories are non-unitary coset models $(A_1)_1\times (A_1)_{L}/(A_1)_{L+1}$ at the fractional level $L=\frac{2}{2N+1}-2$, which appear in the study of the 4d/2d correspondence of ${\cal N}=2$ superconformal field theories. 
Based on the WKB analysis, we  clarify the relation between the Y-functions and the quantum periods and  study the exact Bohr-Sommerfeld quantization condition for the quantum periods.
We also discuss the quantum spectral curves for the D and E type theories.

 \end{abstract}
\end{titlepage}
\baselineskip=0.7cm
\section{Introduction}
The Argyres-Douglas (AD) theory is a strongly coupled ${\cal N}=2$ superconformal field theory in four dimensions, where mutually non-local monopoles/dyons become massless \cite{Argyres:1995jj}. The AD theory is realized by degeneration of the Seiberg-Witten (SW) curve of ${\cal N}=2$ gauge theories \cite{Argyres:1995xn,Eguchi:1996vu}. 
More general methods to obtain the AD theories are to use the hypersurface singularities in type IIB superstrings \cite{Wang:2015mra} or the Hitchin systems \cite{Xie:2012hs}.
Recently there has been recognized a remarkable correspondence between 4d ${\cal N}=2$ superconformal field theories and 2d conformal field theories \cite{Beem:2013sza}.
For the AD theories this correspondence is confirmed by comparing  
 the central charges in both theories and calculating 
the superconformal indices and the characters of the 2d chiral algebras
 \cite{Cordova:2015nma,Buican:2015ina}.
It has been conjectured that a class of the AD theories
correspond to non-unitary  
W minimal models \cite{ Wang:2015mra,Xie:2012hs,Xie:2016evu}.
Recently this conjecture has been proved by using the Hitchin system \cite{Fredrickson:2017yka}.

One can study ${\cal N}=2$ theories  in the Nekrasov-Shatashvili limit of the $\Omega$-background
\cite{Nekrasov:2009rc}. The SW curve is quantized  where the deformation parameter plays a role of the Planck constant \cite{Mironov:2009uv}.
In particular Gaiotto studied the quantized SW curve for the $A_1$ and $A_2$-type AD theories and compared the wave functions for the quantum curves with the solutions of  the thermodynamic Bethe Ansatz (TBA) equations \cite{Gaiotto:2014bza}. 
The ADE-type TBA equations and the associated Y-systems were obtained in \cite{Cecotti:2014zga} from the analysis of the wall-crossing of the BPS spectrum \cite{Gaiotto:2008cd}.
For the $A_n$-type AD theory the quantum spectral curve becomes the second order ordinary differential equation with  
the $(n+1)$-th order potential. 
It turns out that the same ODE appears in the context of the ODE/IM correspondence. So it would be interesting to apply the method of the ODE/IM correspondence to the quantum spectral curve of the AD theories.
We will find the integrable model represented by conformal field theory has the same properties 
as the CFT predicted by the 4d/2d correspondence.

The ODE/IM correspondence describes a relation between the spectral analysis approach of ordinary differential equation (ODE) and the functional relations approach to the 2d quantum integrable model (IM) \cite{Dorey:1998pt,Bazhanov:1998wj,Dorey:2007zx}. 
The ordinary differential equations for the integrable models associated with the Bethe equations 
of a classical Lie algebra ${\mathfrak g}$
have been proposed in \cite{Dorey:2006an}. 
For a general Lie algebra ${\mathfrak g}$, it is 
obtained from the conformal limit  of the linear problem for the modified affine Toda field equations associated with $\hat{\mathfrak g}^{\vee}$, while
the full linear problem describes the massive version of the ODE/IM correspondence \cite{Lukyanov:2010rn, Dorey:2012bx,Ito:2013aea,Adamopoulou:2014fca,Ito:2015nla,Masoero:2015lga,Masoero:2015rcz} (see also \cite{Sun:2012xw} for a classical Lie algebra case).
Here $\hat{\mathfrak g}$ denotes the untwisted affine Lie algebra of ${\mathfrak g}$ and
$\hat{\mathfrak g}^{\vee}$ the Langlands dual of $\hat{\mathfrak g}$.
The Stokes coefficients of the solutions of the ODE are represented by the Wronskians.
They satisfy the functional relations called the T-systems and the Y-systems. 
We can identify the associated integrable models from  the TBA equations which are obtained from the Y-system
\cite{Zamolodchikov:1991et}.

In this paper we study the quantum spectral curves of the AD theories of the ADE-type from the viewpoint of the ODE/IM correspondence. 
In particular we discuss the Y-system and the TBA equations for the $A_n$-type AD theories without flavor symmetry.
We find that the integrable model is identified with the non-unitary minimal model which is predicted by the 4d/2d correspondence.

This paper is organized as follows:
In sec.~\ref{sec:QSW-AD}, we  review the quantum deformation of the Seiberg-Witten curve of the AD theory. In sec.~\ref{sec:QSW-ODE/IM}, we study the  quantum spectral curves of the $A_n$-type AD theories via the ODE/IM correspondence. We construct the T-/Y-systems from the quantum spectral curves and find the corresponding two-dimensional conformal field theories. 
We study the explicit relation between the Y-functions and the quantum periods and  the exact Bohr-Sommerfeld quantization condition satisfied by the quantum periods.  
In sec.~\ref{sec:A_n-A_m},  We study the $(A_m, A_n)$-type AD theories and construct the T-/Y-system. 
We also discuss a generalization to the $D_n$ and $E_n$ type AD theories in sec.~\ref{sec:D_n-E}. 
 Sec.~6 contains conclusions and discussion. 
 In the appendix the matrix representations of the E-type Lie algebras are presented.
 
\section{Quantum SW curve for the Argyres-Douglas theory}\label{sec:QSW-AD}
In this section we review the Seiberg Witten curve of the Argyres-Douglas theory \cite{Argyres:1995jj} 
and its quantum deformation. 
We will consider the AD theory obtained by the degeneration of the SW curve of 
${\cal N}=2$ gauge theory with gauge group $G$ of ADE-type \cite{Klemm:1994qs}. The SW curve for the gauge group $G$ of a Lie algebra ${\mathfrak g}$ is obtained as the spectral curve of the periodic Toda lattice associated with the dual of its affine Lie algebra $\hat{\mathfrak g}^{\vee}$ and a representation $R$ of ${\mathfrak g}$ \cite{Martinec:1995by}. 
After the degeneration of the curve, the SW curve becomes (see \cite{Ito:1999cc} for a review)
\begin{align}\label{eq:SWCurve-AD}
 \xi^2=W^{R}_{\mathfrak g}(x,u_1, \ldots, u_n), 
\end{align} 
where $W_{\mathfrak g}^R$ is the superpotential of a variable $x$ and  the Coulomb moduli parameters $u_1,\ldots, u_n$. Here $n$ denotes the rank of ${\mathfrak g}$. 
For example, $W_{A_n}^{\underline{n+1}}$ for the Lie algebra $A_n$ and  the $(n+1)$-dimensional fundamental representation
$\underline{n+1}$ is given by
\begin{align}
W_{A_n}^{\underline{n+1}}&=x^{n+1}-u_1x^{n-1}- \cdots -u_{n-1}x-u_{n}. 
\end{align}
The SW differential for the curve (\ref{eq:SWCurve-AD}) 
is defined by
$\lambda_{SW}=\xi dx$. 
This differential determines a symplectic structure $d\lambda_{SW}=d\xi\wedge dx$ in the $(\xi,x)$-space.

Now we consider the AD theory deformed in the NS limit of  the $\Omega$-background \cite{Nekrasov:2009rc}. In this limit,
the deformed  theory can be described by the 
quantum spectral curve, where $\xi$ becomes a differential operator $\hat{\xi}=-\epsilon\frac{d}{dx}$.
The quantum spectral curve is defined by a differential equation
\begin{align}\label{eq:QSW-A-type}
\left(-\epsilon^2\partial_x^2+W_{\mathfrak g}^{R}(x,u_1,\ldots,u_n)
\right)\psi(x)=0.
\end{align}

Let us consider the $A_n$ type theory which has the  superpotential with $u_1=\cdots=u_{n-1}=0$ and $u_{n}=-v$,
\begin{align}\label{eq:W-A}
W_{A_n}^{\underline{n+1}}(x)=x^{n+1}+v .
\end{align}
Then  (\ref{eq:QSW-A-type}) becomes 
\bee\label{eq:QSW-A_1-A_n}
\left(-\epsilon^2\partial_x^2+x^{n+1}+v\right)\psi(x)=0.
\ee
Rescaling  $x$ and $v$ as
\begin{align}
x=\epsilon^{\frac{2}{n+3}}z, \quad v=-\epsilon^{\frac{2(n+1)}{n+3}}E,
\label{eq:rescale1}
\end{align}
we get the 
differential equation
\bee
\left(-\partial_{z}^{2}+z^{n+1}-E\right)\psi(z) & = & 0\label{eq:ODE-z}.
\ee
This is the ODE appearing in the ODE/IM correspondence \cite{Dorey:1998pt,Bazhanov:1998wj} without
the potential term ${\ell(\ell+1)\over z^2}$, which produces non-trivial monodromy around the origin.

We  can apply this procedure to the spectral curve $\xi^{m+1}=x^{n+1}+v$,  
which corresponds to the  
$(A_m,A_n)$-type AD theories  \cite{Wang:2015mra,Cordova:2015nma}.
Rescaling $x$ and $v$ to $z$ and $E$, respectively, we obtain the  quantum spectral curve
\bee\label{eq:ode-an-am-0}
\left((-1)^m\partial_z^{m+1}+z^{n+1}-E \right)\psi(z)=0.
\ee
This belongs to the special class of  the  $A_m$-type ODE \cite{Dorey:2000ma},
whose solutions have trivial monodromy around the origin.
In this paper, we call the equation (\ref{eq:ode-an-am-0}) the ($A_m,A_n$)-type ODE.
The quantum curve (\ref{eq:ODE-z}) corresponds to the $(A_1,A_n)$-type ODE.

The $(A_m,A_n)$-type ODE and the $(A_n,A_m)$-type ODE
are related to each other
by the Fourier transformation in $(\xi,x)$-space.
This equivalence between the $(A_m,A_n)$ and $(A_n,A_m)$-type AD theories has been observed in  \cite{Cordova:2015nma}.

In the following we will study the $A_n$ type quantum spectral curve using the ODE/IM correspondence.
We investigate the integrable model using the functional relations which are obtained from the ODE.

\section{$A_{n}$-type quantum spectral curve}\label{sec:QSW-ODE/IM}
In this section we study the quantum spectral curve of the $(A_1, A_n)$-type AD theories using the ODE/IM correspondence. After reviewing the ODE/IM correspondence \cite{Dorey:2007zx}, we discuss  the T-/Y-system obtained from the $(A_1, A_n)$-type ODE and the related integrable model. 
Based on the WKB analysis, we study the relation between the Y-functions and the quantum periods, and study the exact Bohr-Sommerfeld quantization condition for the quantum periods.

\subsection{$(A_1, A_n)$-type ODE} 
We study the ODE/IM correspondence for the quantum spectral curve (\ref{eq:ODE-z}).  
Let  $y(z,E)$ be a solution of 
the differential equation (\ref{eq:ODE-z}), which decays along the real positive axis as
\begin{align}\label{eq:subsol1}
y(z,E) & \sim\frac{z^{-\frac{n+1}{4}}}{\sqrt{2i}}\exp\left(-\frac{z^{\frac{n+3}{2}}}{\frac{n+3}{2}}\right), \quad |z|\rightarrow \infty.
\end{align}
This is called the subdominant solution, which is uniquely defined  in the sector ${\cal S}_{0}: |\mbox{arg}(z)|<\frac{\pi}{n+3}$. Since the differential equation (\ref{eq:ODE-z}) is invariant under the rescaling $(z,E)\rightarrow (\omega z,\omega^{-2}E)$ with
$\omega=e^{\frac{2\pi i}{n+3}}$, one can define the Symanzik rotation of $y(z,E)$ by
\begin{align}
y_{k}(z,E)=\omega^{\frac{k}{2}} y(\omega^{-k}z,\omega^{2k}E),
\label{eq:sym1}
\end{align}
for an integer $k$.
$y_k(z,E)$ is the subdominant solution in the  
 sector ${\cal S}_{k}: |\mbox{arg}(z)-\frac{2k\pi}{n+3}|<\frac{\pi}{n+3}$.  

There are $n+3$ sectors ${\cal S}_0,\ldots, {\cal S}_{n+2}$ which cover the complex $z$-plane.
Since both $y_k$ and $y_{k+n+3}$ are 
 subdominant in the same sector ${\cal S}_k$,
$y_{k+n+3}(z)$ is proportional to $y_k(e^{-2\pi i}z)$.
Moreover the solutions of (\ref{eq:ODE-z}) have trivial monodromy around $z=0$. We find 
\bee\label{eq:monodromy}
y_{k+n+3}(z)\propto y_k(e^{-2\pi i}z)=y_k(z).
\ee
This relation is important to determine the boundary conditions of the T-/Y-system.

We define the Wronskian
\begin{align}
W_{k,i}:=y_ky_i'-y_k'y_i
\label{eq:wronskian1}
\end{align}
for the solutions $y_k$ and $y_i$. $W_{k,i}$ is independent of $z$ and is regarded as a function of $E$. We can choose $\{y_0,y_1\}$ as a  basis of the solutions of the ODE because $W_{0,1}\neq 0$.
We have determined the normalization factor in (\ref{eq:subsol1}) such that 
 $W_{0,1}=1$. 
 From the Symanzik rotation (\ref{eq:sym1}), we have $W_{k,i}(\omega^2 E)=W_{k+1,i+1}(E)$,
 which leads to $W_{k,k+1}=1$.
 
 We expand $y_k$ in the basis $\{ y_0,y_1\}$ as
\bee
y_k=-{W_{1,k}} y_0+{W_{0,k}} y_1.
\ee
We define the T-functions $T_s(E)$ by
\begin{eqnarray}
T_{s}(E)=W_{0,s+1}^{[-(s+1)]},
\end{eqnarray}
where we have used the notation
$f^{[j]}(E):=f(\omega^{j}E)$. 
From the Pl\"ucker relation of the $2\times 2$ determinants
\begin{align}
W_{k_{1},k_{2}}W_{k_{3},k_{4}}=W_{k_{1},k_{4}}W_{k_{3},k_{2}}+W_{k_{3},k_{1}}W_{k_{4},k_{2}},
\end{align}
we obtain the T-system:
\begin{eqnarray}\label{eq:T-system-A_1-A_n}
T_{s}^{[+1]}T_{s}^{[-1]}=T_{s+1}T_{s-1}+1.
\end{eqnarray}
It is obvious that $T_{-1}=0$ and $T_0=1$ from their definitions. From  (\ref{eq:monodromy}) we find the boundary conditions: $T_{n+1}=const.$ and $T_{n+2}=0$.
We introduce the Y-functions by 
\begin{align}
Y_s:=T_{s-1}T_{s+1}.
\label{eq:yfunction1}
\end{align}
The Y-functions satisfy the Y-system:
\bee\label{eq:Y-A_1-A_n}
Y_{s}^{[+1]}Y_{s}^{[-1]}=(1+Y_{s-1})(1+Y_{s+1}).
\ee
The boundary conditions for the Y-functions are $Y_0=Y_{n+1}=0$. Then (\ref{eq:Y-A_1-A_n}) becomes the $A_{n}-$type Y-system. 

\subsection{TBA equation and the central charge}
From the Y-system (\ref{eq:Y-A_1-A_n}), we will obtain the Thermodynamic Bethe Ansatz (TBA) equations \cite{Zamolodchikov:1991et,Dorey:2007zx}. We introduce the parameter $\theta$ by $e^{\theta/\mu}=E$ where $\mu=\frac{n+3}{2(n+1)}$. We also introduce the pseudo-energy by
\begin{align}
\epsilon_s(\theta)=\log Y_s(\theta),~~s=1,\ldots,n.
\end{align}
It is known that the pseudo-energy take the form  $m_k L e^\theta$ for large $\theta$ \cite{Sibuya,Dorey:2007zx}. The coefficients $m_s L$ of $e^{\theta}$ are determined by the Y-system as
\bee
\epsilon_s(\theta)\sim m_sLe^\theta,~~m_sL=\frac{b_0}{2}\sin(\frac{\pi s}{n+1}),~~s=1,\ldots,n.
\label{eq:asymy1}
\ee
Here $b_0$ is given by \cite{Dorey:2007zx}
\bee
\frac{b_0}{2}=-2\frac{\Gamma(-\mu)\Gamma(\mu+\frac{1}{2})}{\sqrt{\pi}}\cos(\frac{\pi}{n+1}).
\ee
With this asymptotics (\ref{eq:asymy1}), one can convert the Y-system to the TBA equations
\bee\label{eq:TBA-A_1-A_n}
\epsilon_{s}(\theta)=m_{s}Le^{\theta}-\frac{1}{2\pi}
\sum_{s'=1}^{n}\int_{-\infty}^{\infty}d\theta'\phi_{ss'}(\theta-\theta')\log(1+e^{-\epsilon_{s'}})(\theta'),~~s=1,\ldots,n
\ee
where  the integration kernel $\phi_{ss'}(\theta)$ is obtained as the Fourier transformation of 
\begin{align}
\tilde{\phi}_{ss'}(p)&=-2\pi \sum_{l=1}^{n}
(2\delta_{sl}\cosh \left({\pi p\over h}\right)-G_{sl})^{-1}G_{ls'}.
\end{align}
Here $h=n+1$ is the Coxeter number of $A_n$ and $G_{sl}$ is the incidence matrix of $A_n$.

From the TBA equation (\ref{eq:TBA-A_1-A_n}), we can evaluate  the ground state energy as
\begin{eqnarray}
E_0(L)=-\frac{\pi c_{eff}}{6L},
\end{eqnarray}
where 
\bee
c_{eff}=\frac{3}{2\pi^{2}}\sum_{s=1}^{n}\int_{-\infty}^{\infty}d\theta m_{s}Le^{\theta}\log(1+e^{-\epsilon_{s}(\theta)})d\theta.
\ee
In the UV limit, the effective central charge is evaluated as
\bee\label{eq:eff-c-A_n}
c_{eff}=\frac{n}{n+3}.
\ee
For even $n$, due to the double counting of the nodes by the identification  $s\leftrightarrow (n+1)-s$ of the $A_n$ Dynkin diagram, there is a factor of two difference from the central charge of the parafermionic CFT \cite{Dorey:2007zx}.
The  central charge (\ref{eq:eff-c-A_n}) is also obtained from the NLIE approach \cite{Dorey:2007zx}.

The effective central charge (\ref{eq:eff-c-A_n}) corresponds to that of  the non-unitary CFT with 
 the central charge $c= 1- \frac{3 (n+1)^2}{n+3}$ and the primary field with the lowest conformal dimension $\Delta_{min}=\frac{1-(n+1)^2}{8 (n+3)}$, where
the effective central charge is defined by $c_{eff}:=c-24\Delta_{min}$.
For even $n$, this non-unitary CFT is identified with the minimal model ${\cal M}_{2,n+3}$ 
\cite{Bazhanov:1994ft}. 
This minimal model is also realized by the coset CFT with fractional  level
\cite{Mathieu:1990dy}
\bee\label{eq:IM-A_1-A_n}
\frac{(A_1)_{1}\times(A_1)_{L}}{(A_1)_{L+1}},\quad L=\frac{2}{n+1}-2.
\ee
The correspondence between (\ref{eq:ODE-z}) and (\ref{eq:IM-A_1-A_n}) was also confirmed by using other functional relations such as Baxter's T-Q relations and the quantum Wronskians \cite{Bazhanov:1994ft, Bazhanov:1996dr}.

Let us compare this  2d integrable model (\ref{eq:IM-A_1-A_n}) obtained from the ODE/IM correspondence with the 2d CFT obtained by the 4d/2d correspondence.
 The chiral algebra corresponding to AD theories (with no flavor symmetry) engineered from the M5 branes have been classified in \cite{Xie:2016evu}. 
Here we focus on the AD theory without flavor symmetry.
For the $A_n$-type AD theories where $n$ is even, the corresponding 2d CFT is 
the coset CFT  (\ref{eq:IM-A_1-A_n}). Moreover in  \cite{Cecotti:2014zga} the TBA equations have been derived from the BPS spectrum \cite{Gaiotto:2008cd}.
For the $A_n$-type AD theories, the Y-system is exactly the same as (\ref{eq:Y-A_1-A_n}).

\subsection{Y-functions and  quantum periods}
We now study the WKB solution of 
the quantum spectral curve (\ref{eq:QSW-A-type}). The WKB solution $\psi(x)$ of (\ref{eq:QSW-A-type}) takes the form
\bee\label{eq:exact-psi}
\psi(x)=c(\epsilon) \exp\left(\pm\frac{1}{\epsilon}\int^xP(x')dx'\right),
\ee
where $c(\epsilon)$ is a constant and $P(x)$ satisfies
\begin{align}
P(x)^2\pm \epsilon P'(x)-P_0(x)^2=0,
\end{align}
with $P_0(x):=(W^R_{\mathfrak{g}}(x))^{1\over2}$.
Substituting the expansion $P(x)=\sum_{n=0}^{\infty}\epsilon^n P_n(x)$,  
$P_n(x)$ can be determined recursively.
It is convenient to decompose $P(x)$ into two parts:
\bee
P(x)=P_{even}(x)+P_{odd}(x),
\ee
where $P_{even}(x)=\sum_{m=0}^\infty\epsilon^{2m}P_{2m}(x)$ and $P_{odd}(x)=\sum_{m=0}^\infty\epsilon^{2m+1}P_{2m+1}(x)$. $P_{odd}(x)$ can be written in terms of 
$P_{even}(x)$ as
\bee
P_{odd}(x)=\mp \frac{\epsilon}{2}\partial_x\log P_{even}(x).
\ee

Now we consider the $(A_1,A_n)$-type ODE (\ref{eq:QSW-A_1-A_n}) where $P_0(x)=(x^{n+1}+v)^{1/2}$. 
The subdominant solutions $y_k(z,E)$ of (\ref{eq:ODE-z}) are regarded as the function of $x$ and are parametrized by $v$ and $\epsilon$, which are denoted  by
$y_k(x,v,\epsilon)$. In this parametrization of the solution, the Symanzik rotation $(z,E)\rightarrow (\omega z,\omega^{-2}E)$ is also realized by  $\epsilon\rightarrow -\epsilon$. The WKB expansion of $y_k(x,v,\epsilon)$ is obtained  as
\bee\label{eq:y-exact-WKB}
y_k(x,v,\epsilon)=(-1)^{\frac{k}{2}}c(\epsilon)\exp\left(\frac{{\delta}_k}{\epsilon}\int^{x}_{x_k}P(x')dx'\right),
\ee
where $c(\epsilon)={1\over \sqrt{2i}}\epsilon^{\frac{(n+1)}{2(n+3)}}$.
Here we have fixed the initial point $x_k$ of the contour integration.
${\delta}_k=\pm (-1)^k$ is the sign factor, where
the minus sign is taken  when the integration variable $x'$ belongs to the sheet $\xi=+P_0(x)$ 
 of the Riemann surface $\xi^2=P_0^2(x)$.
The factor $(-1)^k$ comes from the Symanzik rotation. Equivalently we can choose the plus sign when $x'$ belongs to the sheet $\xi=-P_0(x)$.

We will evaluate the Wronskian $W_{k,i}$ for the solutions $y_k$ and $y_i$. Substituting the WKB solution (\ref{eq:y-exact-WKB}) into the Wronskian (\ref{eq:wronskian1}), we get
\bee\label{eq:Wki}
W_{k,i}=i(-1)^{\frac{k+i}{2}}\delta_k\exp\left(\frac{\delta_k}{\epsilon}\int^{x_i}_{x_k}P_{even}(x')dx'+\frac{1}{2}[\log P_{even}(x_i)+\log P_{even}(x_k)]\right).
\ee
Here we have chosen the sheets such  that  the explicit $x$ dependence  does not appear in the expression.
This can be done when the condition $\delta_k=-\delta_i $ holds.

\begin{figure}[htpb]
\begin{center}
\resizebox{70mm}{!}{\includegraphics{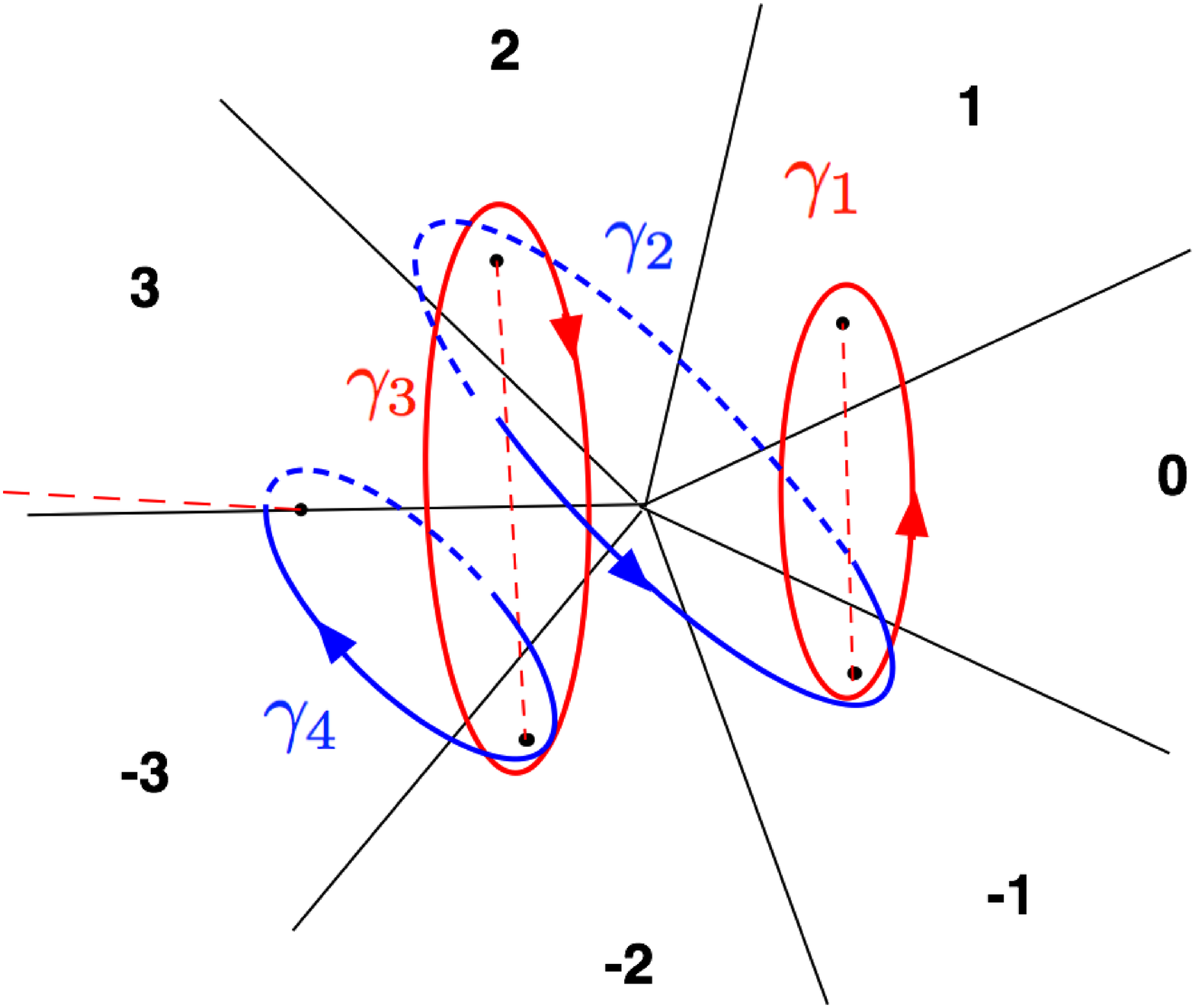}}
\end{center}
\caption{The integration contours $\gamma_s$ for the Y-functions for the $A_4$ case. The contour is defined on the Riemann surface $\xi^2=x^5+v$. The numbers stand for the number of sectors. The black dots stand for the zeros of $x^5+v$. The red dot lines are the branch cuts.   
}
\label{fig:A4}
\end{figure}

We then define the Y-function (\ref{eq:yfunction1}) 
\bee\label{eq:new-Y}
Y_{2j}:=\frac{W_{-j,j}W_{-j-1,j+1}}{W_{-j-1,-j}W_{j,j+1}},~~Y_{2j+1}=\left[\frac{W_{-j-1,j}W_{-j-2,j+1}}{W_{-j-2,-j-1}W_{j,j+1}}\right]^{[+1]}.
\ee
Note that the present Y-functions have also appeared in the study of minimal surface in $AdS_3$ spacetime \cite{Alday:2010vh}.
In (\ref{eq:new-Y}) the normalization factors of $Y_s$ are different from
those in (\ref{eq:yfunction1}), where they are simply set to be 1.
From (\ref{eq:Wki}), the Y-functions (\ref{eq:new-Y}) are expressed  by the quantum periods \cite{Mironov:2009uv}
\begin{align}\label{eq:Y-asym-eps}
 Y_s=\exp\left(i^{\frac{(-1)^{s}-1}{2}}\frac{1}{\epsilon}
 \oint_{\gamma_{s}} P_{even}(t)dt
\right),~~s=1,\ldots,n,
\end{align}
where $\gamma_{s}$ is the cycles defined as follows.
We label the zeros of $x^{n+1}+v$ (for even $n$) as $p_m=v^{\frac{1}{n+1}}\eta^{2m+1}$ $(m=0,1,\ldots,{n\over2})$ where $\eta=\exp({\pi i\over n+1})$.
Then $\gamma_{2j+1}$ are the cycles that enclose the cut between $p_{-(j+1)}$ and $p_{j}$ (counter)clockwise for odd (even) $j$. $\gamma_{2j}$ are the cycles enclosing $p_{-j}$ and $p_{j}$ (counter)clockwise for even (odd) $j$. 
In Fig.~\ref{fig:A4}, we describe the  contours for the $A_4$ case.

We can evaluate $\log Y_s$ as the power series in $\epsilon$.
At the leading order in $\epsilon$, we have 
\begin{align}
\log Y_{s}\sim i^{\frac{(-1)^{s}-1}{2}}\frac{\Pi_{\gamma_{s}}}{\epsilon},
\end{align}
where 
\begin{align}
\Pi_{\gamma_{2j+1}}=2(-1)^{j+1} I(p_{-j-1},p_{j}),\quad 
\Pi_{\gamma_{2j}}=2(-1)^jI(p_{-j},p_j)
\end{align}
and
\bee
I(p_m,p_{m'}):=\int_{p_m}^{p_{m'}}P_0(t)dt=v^{\frac{1}{n+1}+\frac{1}{2}}\int_{\eta^{2m+1}}^{\eta^{2m'+1}}\sqrt{t^{n+1}+1}dt.
\ee
The integral is evaluated as
\begin{align}
I(p_m,p_{m'})=-\epsilon (-E)^\mu\frac{\Gamma(\mu+\frac{1}{2})\Gamma(-\mu)}{\sqrt{\pi}}\cos(\frac{\pi}{n+1})\eta^{m+m'+1}i\sin\left({(m-m')\pi\over n+1}\right).
\end{align}
Therefore the asymptotic behavior of $\log Y_{s}$ is given by
\begin{align}\label{eq:period-Int-Y}
\log Y_{s}&\sim -
(-E)^\mu\frac{\Gamma(\mu+\frac{1}{2})\Gamma(-\mu)}{\sqrt{\pi}}\cos(\frac{\pi}{n+1}) 2i^{s-1}\eta^{\frac{1+(-1)^s}{2}} \sin({s\pi\over n+1}).
\end{align}
This agrees with  (\ref{eq:asymy1}) 
up to a phase factor of  $i^{s-1}\eta^{\frac{1+(-1)^s}{2}}$, which
comes from the argument of  the integration contour in the complex plane.

\subsection{Exact Bohr-Sommerfeld quantization condition}
We have seen that the Y-function is related to the quantum period. For some special values of $E$,  the quantum periods satisfy the Bohr-Sommerfeld quantization condition. For a fixed $k$, we consider the boundary condition for the solution of 
(\ref{eq:ODE-z})
such that the solution decays exponentially at large $|z|$ in 
the sectors ${\cal S}_0$ and ${\cal S}_{k+1}$: 
\bee \label{eq:bc1}
\psi(z,E)\propto y_0,~~z\in {\cal S}_0,~~\label{eq:B.C.psi-01},~~
\psi(z,E)\propto y_{k+1},~~z\in {\cal S}_{k+1}.
\ee
Since $\psi=y_0$ in ${\cal S}_0$ can be expressed  in terms of the basis $y_{k+1}$ and $y_{k+2}$:
\bee
y_0=\frac{W_{0,k+2}}{W_{k+1,k+2}}y_{k+1}+ \frac{W_{k+1,0}}{W_{k+1,k+2}}y_{k+2}.
\ee
Then the boundary condition (\ref{eq:bc1}) implies that 
\bee
W_{0,k+1}(E)=0,
\label{eq:bs1}
\ee
which can be regarded as the condition on the parameter $E$.
This equation can be written in the form
\bee\label{eq:quant-cond-j}
\log Y_k(\theta+k\frac{\pi i}{n+1})=(2I+1)\pi i
\ee
for an integer $I$ and a fixed $k$. 
At large $\theta$, from (\ref{eq:asymy1}), this condition 
is expressed as
\bee\label{eq:BS-cond-1th}
m_kL e^{\theta+\frac{k\pi i}{n+1}}\sim (2I+1)\pi i, 
\ee
where $I$ is a large integer.
This is the quantization condition for $\theta$. For $k=1$, we note that the condition (\ref{eq:BS-cond-1th}) on $E$ coincides with the one obtained in \cite{Sibuya} which is based on the zeros of $W_{0,2}(E)$ at large $E$. From (\ref{eq:Y-asym-eps}), (\ref{eq:BS-cond-1th}) is written in the form
\bee
&(-i\eta)^{k-\frac{1+(-1)^{k}}{2}}\frac{1}{\epsilon}\oint_{\gamma_{k}}P_0(t) dt\sim(2I+1)\pi i.
\ee
This is the Bohr-Sommerfeld quantization condition for the classical period.

We have argued that  the appropriate boundary condition of the solution of the quantum spectral curve leads to the condition on the argument of the Y-function and the quantum deformed period. The convolution term in TBA equation provides the higher order corrections to the Bohr-Sommerfeld quantization condition.

\section{$(A_m,A_n)$-type quantum spectral curve}\label{sec:A_n-A_m}
In this section, we study 
the $(m+1)$-th order ODE (\ref{eq:ode-an-am-0}), which is associated with the Lie
algebra $A_m$ \cite{Dorey:2006an}. 
We introduce a parameter $M$ by $(m+1)M=n+1$ in (\ref{eq:ode-an-am-0}). Then the ODE  (\ref{eq:ode-an-am-0}) becomes
\begin{eqnarray}\label{eq:A_n-A_m}
\left((-1)^m\frac{d^{m+1}}{dz^{m+1}}+z^{(m+1)M}-E\right)\psi(z,E)=0.~~~
\end{eqnarray}
Let $y(z,E)$ be the solution of (\ref{eq:A_n-A_m}), which is subdominant in the sector ${\cal S}_0: |\mbox{ arg}(z)|\leq \frac{\pi}{(m+1)(M+1)}$.  
$y(z,E)$ has asymptotic behavior along the positive real axis as 
\begin{eqnarray}
y(z,E)\sim\frac{z^{-mM/2}}{i^{m/2}\sqrt{m+1}}\exp\left(-\frac{z^{M+1}}{M+1}\right),\quad |z|\to\infty. 
\end{eqnarray}
The ODE (\ref{eq:A_n-A_m}) is invariant under the Symanzik rotation $(z,E)\to (\omega z, \omega^{(m+1)M}E)$ with $\omega=e^{\frac{2\pi i}{n+m+2}}$.
Then for integer $k$ 
\begin{eqnarray}
y_{k}(z,E)=\omega^{mk/2}y(\omega^{-k}z,\omega^{-(m+1)Mk}E)
\end{eqnarray}
is  the solution which is subdominant in the sector ${\cal S}_k: |\mbox{ arg}(z)-\frac{2\pi k}{(m+1)(M+1)}|\leq \frac{\pi}{(m+1)(M+1)}$. There are $n+m+2$ sectors ${\cal S}_0,\ldots, {\cal S}_{n+m+1}$ which cover the $z$-complex plane.
Since  the solutions have trivial monodromy around $z=0$, we have
\bee
y_{-m+1}(z)\propto y_{-m+1+m+n+2}(z)=y_{n+3}(z).\label{eq:no-sing-A_n-A_m}
\ee
The Wronskians of the solutions $y_{i_1},\ldots, y_{i_{m+1}}$ is defined by
\begin{align}\label{eq:Wronskian}
W[y_{i_1},y_{i_2},\ldots, y_{i_{m+1}}]=\det\left(\begin{array}{cccc}
y_{i_{1}} & y_{i_{2}} & \cdots & y_{i_{m+1}}\\
\partial_{z}y_{i_{1}} & \partial_{z}y_{i_{2}} & \cdots & \partial_{z}y_{i_{m+1}}\\
\cdots & \cdots &  & \cdots\\
\partial_{z}^{m+1}y_{i_{1}} & \partial_{z}^{m+1}y_{i_{2}} & \cdots & \partial_{z}^{m+1}y_{i_{m+1}}
\end{array}\right).~~
\end{align}
The Wronskian (\ref{eq:Wronskian}) is independent of $z$ and is regarded as a function of $E$. Defining $f^{[ j]}(E):=f(\omega^{- j\frac{(m+1)M}{2}}E)$, 
the Wronskian satisfies the relation
$$W[y_{i_1},y_{i_2},\ldots, y_{i_{m+1}}]^{[+2]}=W[y_{i_1+1},y_{i_2+1},\ldots, y_{i_{m+1}+1}].$$
Then we introduce the T-functions $T_{s,l}$ $(1\leq s\leq m,l\in {\bf Z})$ by
\begin{align}
T_{2\hat{m},l}&=W[{y_{-m+\hat{m}+1},y_{-m+\hat{m}+2},\ldots, y_{-\hat{m}+1}, y_{l+2-\hat{m}}, y_{l+3-\hat{m}},\ldots , y_{l+\hat{m}+1}}]^{[-l-1]},
\no\\
T_{2\hat{m}+1,l}&=W[{y_{-m+\hat{m}+1},y_{-m+\hat{m}+2},\ldots, y_{-\hat{m}}, y_{l+1-\hat{m}},y_{l+2-\hat{m}},\ldots, y_{l+\hat{m}+1}}]^{[-l]}.
\end{align}
From the Pl\"ucker relation for the determinants of $(m+1)\times (m+1)$ matrices:
\begin{align}
	0=W[y_{j_{1}},y_{j_{2}},\ldots,y_{j_{k-2}},y_{j_{k-1}},y_{i_{1}}]W[ y_{j_{1}},y_{j_{2}},\ldots,y_{j_{k-2}},y_{i_{2}},y_{i_{3}}]\no\\
    -W[ y_{j_{1}},y_{j_{2}},\ldots ,y_{j_{k-2}},y_{j_{k-1}},y_{i_{2}}]W[ y_{j_{1}},y_{j_{2}},\ldots ,y_{j_{k-2}},y_{i_{1}},y_{i_{3}}]\label{eq:Pluker}\\
    +W[y_{j_{1}},y_{j_{2}},\ldots ,y_{j_{k-2}},y_{j_{k-1}},y_{i_{3}}]W[ y_{j_{1}},y_{j_{2}},\ldots,y_{j_{k-2}},y_{i_{1}},y_{i_{2}}]\no,
\end{align}
we find that  the $T_{s,l}$ satisfy the functional relations:
\begin{align}\label{eq:T-sys-A_r}
	&T_{s,l}^{[+1]}T_{s,l}^{[-1]}=T_{s,l+1}T_{s,l-1}+T_{s-1,l}T_{s+1,l},\quad s=1,\ldots,m.
\end{align}
Here $T_{0,l}=T_{m+1,l}=1$ and  
\bee
T_{s,-1}=0, \quad T_{s,0}=1,\quad s=1,\ldots,m.
\ee
The condition (\ref{eq:no-sing-A_n-A_m}) implies that  $T_{s,n+2}=0$ and $T_{s,n+1}$ is a constant. We then get the $(A_m,A_{n})$-type T-system \cite{Ravanini:1992fi}. 
We also introduce the Y-function $Y_{s,l}$ by
\begin{eqnarray}
	Y_{s,l}=\frac{T_{s-1,l}T_{s+1,l}}{T_{s,l+1}T_{s,l-1}}.
\end{eqnarray} 
Then from (\ref{eq:T-sys-A_r}) the Y-functions are shown to satisfy
\begin{eqnarray}\label{AmAn-Y-sys}
	Y_{s,l}^{[+1]}Y_{s,l}^{[-1]}=\frac{(1+Y_{s-1,l})}{(1+Y_{s,l-1}^{-1})}\frac{(1+Y_{s+1,l})}{(1+Y_{s,l+1}^{-1})}.
\end{eqnarray}
From the boundary conditions for the T-functions we obtain
\begin{align}
Y_{0,l}&=0,~Y_{m+1,l}=0,~~~l=1,\ldots n,\\
Y_{s,0}&=\infty,~Y_{s,n+1}=\infty,\quad s=1,\ldots,m.
\end{align}
Using these boundary conditions, (\ref{AmAn-Y-sys}) becomes the $(A_m,A_{n})$-type Y-system \cite{Ravanini:1992fi}. 
From the WKB analysis of the solutions, we may calculate  the asymptotic behavior of the Y-functions at large $\theta$,
which leads to the TBA equations. Since the period integral is rather difficult to evaluate, we left the detailed analysis of the TBA equations for future work. 

The integrable model related to the $(A_m,A_n)$-type ODE has been studied in \cite{Dorey:2006an,Dorey:2007ti} based on the NLIE approach. The ODE (\ref{eq:A_n-A_m}) has been proposed to correspond to the non-unitary minimal model:
\begin{eqnarray}\label{eq:IM-A_n-A_a}
\frac{(A_m)_{L}\times(A_m)_{1}}{(A_m)_{L+1}},
\end{eqnarray}
with the fractional level $L=\frac{1}{M}-(m+1)$.
This is the non-unitary W-minimal model $WA_m(p,q)$ for a pair of coprime integers $p=m+1$ and $q=m+n+2$. 
The  effective central charge of this CFT is given by \cite{Dunning:2002cu}
\bee
c_{eff}=m\left(1-\frac{(m+1)(m+2)}{pq}\right)=\frac{mn}{m+n+2}.
\ee

This 2d CFT corresponds to the $(A_{m}, A_{n})$-AD theory without flavor symmetry as predicted by the 4d/2d correspondence \cite{Xie:2016evu}.

\section{Quantum spectral curves for $D_n$ and $E_n$ cases}\label{sec:D_n-E}

In this section we  will discuss the quantum spectral curve of the $({\mathfrak g}, A_m)$-types, where ${\mathfrak g}$ is a simply-laced Lie algebra. 
We propose that the $({\mathfrak g}, A_m)$-type quantum spectral curve is given by the ODE associated with the  conformal limit of the modified affine Toda field equations associated for the affine Lie algebra $\hat{\mathfrak g}^{\vee}$.
For a classical  Lie  algebra ${\mathfrak g}$ the ODEs have been constructed by 
\cite{Dorey:2006an,Sun:2012xw,Ito:2013aea}. Let us consider the  
linear problem  for the modified affine Toda field equation for $\hat{\mathfrak g}$ of a simply laced Lie algebra ${\mathfrak g}$
of rank $n$.
In the conformal 
limit \cite{Lukyanov:2010rn,Ito:2013aea,Ito:2015nla}, it takes the form:
\bee\label{eq:linear problem}
(\partial_z+A(z))\Psi(z)=0,
\ee
where  $A(z)$ is the connection  defined by
\bee
A(z)=p(z)\sqrt{n_0^\vee}E_{\alpha_0}+\sum^n_{i=1}\sqrt{n_i^\vee}E_{\alpha_i}. \label{eq:flat-conn}
\ee
Here $\alpha_1,\ldots, \alpha_n$ are simple roots of ${\mathfrak g}$  and $\alpha_0:=-\theta$
where $\theta$ denotes the highest root of ${\mathfrak g}$. 
$E_{\alpha_i}$ ($i=0,\ldots n$) are the Chevallery generators of $\hat{\mathfrak g}$. 
$n_i^{\vee}$ are the dual Coxeter labels which are defined by $\theta^{\vee}=\sum_{i=1}^{n}n_i^{\vee}\alpha^{\vee}_i$ and 
$n_0^{\vee}=1$. $\alpha^{\vee}:=2\alpha/\alpha^2$ denotes a coroot of $\alpha$, which is normalized as $\alpha^2=2$ for a simply-laced Lie algebra.
$p(z)$ is  a holomorphic function of $z$ and it is taken to be a polynomial of $z$. For the $({\mathfrak g},A_m)$-type AD theory, $p(z)$ is chosen as $p(z)=z^{m+1}-E$. The linear problem (\ref{eq:linear problem}) can be written in the form of the (pseudo-) ODE for the first component of $\Psi(z)$. 

For the $A_n$-type Lie algebra, we take  the $(n+1)-$dimensional fundamental representation $E_{\alpha_i}=E_{i,i+1}$ ($i=1,\ldots,n$), $E_{\alpha_0}=E_{n+1,1}$, 
where $E_{ij}$ denotes the matrix with the elements $(E_{ij})_{ab}=\delta_{ia}\delta_{jb}$. 
The ODE for the top component $\psi_1(z)$ of $\Psi(z)$ is given by
\bee\label{eq:A_n-p}
\left((-1)^n\partial_z^{n+1}+p(z)\right)\psi_{1}=0.
\ee
Choosing $p(x)=z^{m+1}-E$, one obtains the ($A_n,A_m$)-type quantum spectral curve.
 
\subsection{$D_n$}
For the $D_n$-type Lie algebra, we construct the quantum spectral curve by using the $2n\times 2n$ matrix representation:
$E_{\alpha_i}=E_{i,i+1}+E_{2n-i,2n+1-i}$ ($i=1,\ldots,n-1$), $E_{\alpha_n}=E_{n-1,n+1}+E_{n,n+2}$ and $E_{\alpha_0}=E_{2n-1,1}+E_{2n,2}$. The linear  problem (\ref{eq:linear problem}) becomes the ODE for  the first component $\psi(z)$ of $\Psi(z)$ as
\begin{eqnarray}\label{eq:D_n}
\left(\frac{d^{2n-1}}{dz^{2n-1}}-2^{n-1}\sqrt{p(z)}{d\over dz}\sqrt{p(z)}\right)\psi(z)=0.
\end{eqnarray}
This is a special type of the $D_n$-type ODE, whose solutions have trivial monodromy around $z=0$ \cite{Dorey:2006an}. 
Let us consider the case  
 $p(z)=2^{1-n}(z^2-E)$ as an example. Then (\ref{eq:D_n}) becomes
\begin{eqnarray}
\left(\frac{d^{2n-1}}{dz^{2n-1}}-\sqrt{z^2-E}\frac{d}{dz}\sqrt{z^2-E}\right)\psi(z)=0.
\label{eq:odedn}
\end{eqnarray}
The differential equation has a $(2n-1)$ dimensional basis of the solutions.
We start with the  solution $y(z,E)$ 
 decaying along the  positive real axis. 
From the WKB analysis, it is found to be
\bee
y(z,E)\sim {\cal C} z^{-1}\exp\left(-\frac{z^{M+1}}{M+1}\right),\quad |z|\rightarrow \infty,
\ee
 where $M=\frac{1}{n+1}$ and ${\cal C}$ is a constant. $y(z,E)$ is the subdominant solution  in the sector ${\cal S}_0:|\mbox{arg}(z)|<\frac{\pi}{2n}$. The ODE (\ref{eq:odedn}) is invariant under the rotation $(z,E)\to (\omega z,\omega^{2}E)$ with $\omega=e^{\frac{2\pi i}{2n}}$.
We then define the Symanzik rotated solution $y_k$ ($k\in{\bf Z}$) by
\begin{eqnarray}
y_{k}(z,E)=\omega^{ak}y_{0}(\omega^{-k}z,\omega^{-2k}E),
\end{eqnarray}
where $a$ is a constant to be determined in the following. $y_k(x,E)$ is subdominant in the sector ${\cal S}_k: |\mbox{arg}(z)-\frac{2\pi k}{2n}|<\frac{\pi}{2n}$. We thus have obtained the set of solutions  $\{y_0,y_1,\ldots,y_{2n-2}\}$.
However these do not provide the basis of the solutions since their Wronskian is shown to be zero.
We find the other solution independent of $y_0,\ldots, y_{2n-3}$ whose asymptotic form is given by
\begin{align}
\psi_0(z)&=z^{-1}\exp(\frac{E}{2}z^{-2}+\frac{E^2}{4}z^{-4}+\cdots ), \quad |z|\rightarrow \infty.
\end{align}
The solution is a decaying solution  at $|z|\rightarrow \infty$ and  is invariant under the Symanzik rotation. We can determine the constants ${\cal C}$ and $a$ such that their Wronskian is normalized as
\begin{eqnarray}\label{eq:wronskian3}
W[y_{i_{1}},y_{i_{1}+1},\ldots,y_{i_{1}+2n-3},\psi_{0}]=1
\end{eqnarray}
for any integer $i_1$. 
Since the solutions have trivial monodromy at $z=0$, $y_k$'s satisfy the condition:
\bee\label{eq:period-Dn}
y_{k+2n}(z)=\omega^{2na}y_k(e^{-2\pi i}z)=e^{2\pi ia} y_k(z).
\ee
Let us define $f^{[k]}(E):=f(\omega^{-k}E)$.
Then the Wronskians obey the relation
\bee
W[y_{i_{1}},y_{i_{2}},\ldots,y_{i_{2n-2}},\psi_{0}]^{[+2]}=W[y_{i_{1}+1},y_{i_{2}+1},\ldots,y_{i_{2n-2}+1},\psi_0].
\ee
We introduce the T-functions $\hat{T}_{s,l}$ $(1\leq s\leq 2n-3, l\in \mathbf{Z})$ by
\begin{align}
&\hat{T}_{2\hat{m},l}=W[y_{-r+\hat{m}+1},y_{-r+\hat{m}+2},\ldots, y_{-\hat{m}+1}, y_{l+2-\hat{m}},y_{l+3-\hat{m}},\ldots , y_{l+\hat{m}+1},\psi_0]^{[-l-1]},
\no\\
&\hat{T}_{2\hat{m}+1,l}=W[y_{-r+\hat{m}+1},y_{-r+\hat{m}+2},\ldots, y_{-\hat{m}}, y_{l+1-\hat{m}},y_{l+2-\hat{m}},\ldots , y_{l+\hat{m}+1},\psi_0]^{[-l]},
\label{eq:tdn}
\end{align}
where $r=2n-3$. 
From the Pl\"ucker relation (\ref{eq:Pluker}) for the determinants of $(2n-1)\times (2n-1)$ matrices, we find the functional relations satisfied by the T-functions:
\begin{align}\label{eq:T-sys-A_r-D_r}
	&\hat{T}_{s,l}^{[+1]}\hat{T}_{s,l}^{[-1]}=\hat{T}_{s-1,l}\hat{T}_{s+1,l}+\hat{T}_{s,l+1}\hat{T}_{s,l-1},~s=1,\ldots,2n-3
\end{align}
where $\hat{T}_{0,l}=\hat{T}_{2n-2,l}=1$.  We also have $\hat{T}_{s,-1}=0$ and $\hat{T}_{s,0}=1$ ($s=1,2,\ldots,2n-3$) from the definition (\ref{eq:tdn}). Furthermore (\ref{eq:period-Dn}) leads to $\hat{T}_{s,3}=0$ and $\hat{T}_{s,2}$ is constant ($s=1,2,\ldots,2n-3$) so that
 only $\hat{T}_{s,1}$ is nontrivial. 
 Therefore (\ref{eq:T-sys-A_r-D_r}) becomes the $(A_{2n-3},A_1)$-type T-system. However $\hat{T}_{2n-2-a,1}$ and $\hat{T}_{a,1}$ are not independent. This implies that the T-system is not a generic $(A_{2n-3},A_1)$-type one.
In the following we will investigate the T-system for the
$D_3$ and $D_4$ cases in detail.

For $D_3$, 
from (\ref{eq:wronskian3})
the ${\cal C}$ and $a$ can be fixed as ${\cal C}=\frac{\sqrt{i}}{2}$ and $a=-\frac{1}{4}$.
$\hat{T}_{s,1}$ $(s=1,2,3)$ 
satisfy a $(A_3,A_1)$-type T-system (\ref{eq:T-sys-A_r-D_r}). 
The condition (\ref{eq:period-Dn}) leads to
$\hat{T}_{1,1}^{[+6]}=i\hat{T}_{3,1},~\hat{T}_{2,1}^{[+6]}=-\hat{T}_{2,1}$.
After rescaling $T_s=(-i)^{s/2}\hat{T}_{s,1}$, we obtain the $A_3$-type T-system with the boundary condition
$T_4=-1$.
This can also be seen from the fact that the $D_3$-type ODE  is equivalent to that of $A_3$ \cite{Dorey:2006an}.

We next consider the $D_4$ case where  the constants are fixed as ${\cal C}=\frac{i^{1/6}}{\sqrt{6}}$ and $a=1$. 
$\hat{T}_{s,l}$ satisfy the $(A_5,A_1)$ T-system 
 (\ref{eq:T-sys-A_r-D_r}), which is given by
\begin{eqnarray}
\hat{T}_{1,1}^{[+1]}\hat{T}_{1,1}^{[-1]}&=&\hat{T}_{1,2}+\hat{T}_{2,1},~~
\hat{T}_{2,1}^{[+1]}\hat{T}_{2,1}^{[-1]}=\hat{T}_{2,2}+\hat{T}_{1,1}\hat{T}_{3,1},~~
\hat{T}_{3,1}^{[+1]}\hat{T}_{3,1}^{[-1]}=\hat{T}_{3,2}+\hat{T}_{2,1}\hat{T}_{4,1}\no\\
\hat{T}_{4,1}^{[+1]}\hat{T}_{4,1}^{[-1]}&=&\hat{T}_{4,2}+\hat{T}_{3,1}\hat{T}_{5,1},~~
\hat{T}_{5,1}^{[+1]}\hat{T}_{5,1}^{[-1]}=\hat{T}_{5,2}+\hat{T}_{4,1},\label{eq:T-D_4-A_5}
\end{eqnarray}
where $\hat{T}_{1,2}=\hat{T}_{5,2}=\hat{T}_{3,2}=-1,~\hat{T}_{2,2}=\hat{T}_{4,2}=1$.
(\ref{eq:period-Dn})  leads to  the periodicity conditions:
\bee
\hat{T}_{5,1}^{[-8]}=-\hat{T}_{1,1},~~\hat{T}_{4,1}^{[-8]}=\hat{T}_{2,1},~~\hat{T}_{3,1}^{[-8]}=-\hat{T}_{3,1}.
\ee
We now introduce the new T-function $T_{s,l}$ ($s=1,2,3,4$ and $l\in {\bf Z}$) by
\bee
&&\hat{T}_{1,l}=T_{1,l},~~\hat{T}_{2,l}=T_{2,l},~~
\hat{T}_{3,l}=T_{3,l}T_{4,l},\\
&&\hat{T}_{4,l}-\hat{T}_{2,l}=T_{3,l+1}T_{3,l-1}+T_{4,l+1}T_{4,l-1}.
\ee
The non-trivial T-system exists when $T_{3,-1}=T_{4,-1}=0$ and $T_{3,3}=T_{4,3}=0$. Furthermore, we can rescale the T-functions to set $T_{3,0}=T_{4,0}=1$.
Then the relation of $T_{3,l}^{[+1]}T_{3,l}^{[-1]}$ in (\ref{eq:T-D_4-A_5}) factorizes, which leads to the $(D_4,A_1)$-type T-system: 
\begin{align}
T_{1,l}^{[+]}T_{1,l}^{[-]}&=T_{1,l+1}T_{l-1}+T_{2,l}T_{0,l},~~
T_{2,l}^{[+]}T_{2,l}^{[-]}=T_{2,l+1}+T_{1,l}T_{3,l}T_{4,l},\no\\
T_{3,l}^{[+]}T_{3,l}^{[-]}&=T_{3,l+1}T_{3,l-1}+T_{2,l},~~
T_{4,l}^{[+]}T_{4,l}^{[-]}=T_{4,l+1}T_{4,l-1}+T_{2,l}.\label{eq:D4-T-system}
\end{align}
 
We have redefined the T-functions to construct the  $D_4$-type T-system.  
For $n\geq 5$, it would be possible to construct  the $D_n$-type T-system by 
some redefinitions.
Note that one can also redefine the Q-functions which satisfy the $D_n$-type Bethe ansatz equations from the $A_n$-type Q-fucntions \cite{Dorey:2006an}. 

We now investigate the relation to the $(D_n,A_{m})$-type AD theories. For the $(D_n,A_{m})$-type AD theories without flavor symmetry  $m$ must be even \cite{Xie:2016evu}. 
Let us choose $p(z)=2^{1-n}(z^{m+1}-E)$ in (\ref{eq:D_n}) with even $m$. In \cite{Dorey:2006an,Dorey:2007ti}
the ODE has been proposed to correspond to the non-unitary $WD$ minimal model:
\begin{eqnarray}\label{eq:IM-D_n-A_m}
\frac{(D_n)_{L}\times(D_n)_{1}}{(D_n)_{L+1}}
\end{eqnarray}
with the fractional level $L=\frac{2n-2}{m+1}-(2n-2)$. 
Thus the ODE/IM correspondence for the $D_n$-type ODE leads to the relation between this CFT and the $(D_{n}, A_{m})$-type AD theory, which is 
predicted by the 4d/2d correspondence \cite{Xie:2016evu}.

\subsection{$E_6$, $E_7$ and $E_8$}
We next write down the quantum spectral curves for exceptional type Lie algebras $E_6$ and $E_7$ explicitly.
In the case of $E_6$, we use the 27-dimensional representation to get the linear system $(\partial+A)\Psi=0$
where $\Psi={}^t(\psi_1,\ldots,\psi_{27})$. 
Using a basis of $E_6$ given in the appendix, we can rewrite the linear system into the pseudo-differential equation for the top
component $\psi_1$ of $\Psi$, which is given by
\begin{align}
&6(p'+3p\partial)\partial^{-9}(2p'+3p\partial)\psi_{1}+\frac{1}{\sqrt{3}}p^{(5)}\psi_{1}+\frac{367\sqrt{3}}{24}p^{(4)}\psi_{1}^{(1)}+\frac{21\sqrt{3}}{2}p^{(3)}\psi_{1}^{(2)}\no\\
&+\frac{39\sqrt{3}}{2}p^{(2)}\psi_{1}^{(3)}+\frac{75\sqrt{3}}{4}p^{(1)}\psi_{1}^{(4)}+\frac{15\sqrt{3}}{2}p\psi_{1}^{(5)}-\frac{1}{32\cdot27}\psi_{1}^{(17)}=0.\label{eq:E_6-ODE-p}
\end{align}
Here we introduced a formal inverse differential operator $\partial^{-1}$.

In the case of $E_7$, we use the 56-dimensional representation given in the appendix to get the linear system $(\partial+A)\Psi=0$, where
$\Psi={}^t(\psi_1,\ldots,\psi_{56})$. It reduces to three differential equations
for the components $\psi_1$,$\psi_{7}$ and  $\tilde{\psi}_{16}:=\psi_{16}-{65\over 56\sqrt{3}}\psi_7^{(4)}-{1079\over 9576\sqrt{6}}\psi_1^{(9)}$:
\begin{align}
\pa^{10}\tilde{\psi}_{16}&=-{12\over 133}\sqrt{6}p' \psi_1-{16\over 133}\sqrt{6}p \psi_1' ,
\label{eq:e7-1}\\
(2p+{115\over 48384}\pa^{18})\psi_7&=
\Bigl(
{25\over7}\sqrt{2}p^{(3)}\pa^2+{38\over7}\sqrt{2}p^{(2)}\pa^3+{25\over21}\sqrt{2}p^{(4)}\pa
+{703\over 84\sqrt{2}}p'\pa^4
\nonumber\\
&+{9\over 28\sqrt{2}}p^{(5)}+{17\over7\sqrt{2}}p \pa^5-{11\over 48384\sqrt{2}}\pa^{23}
\Bigr)\psi_1, 
\label{eq:e7-2}
\end{align}
\begin{align}
&\Bigl(
{31\over 912\sqrt{6}}p^{(10)}+{1151\over 2736\sqrt{6}}p^{(9)}\pa
+{1067\over 456\sqrt{6}}p^{(8)}\pa^2+{3509\over 456\sqrt{6}}p^{(7)}\pa^3
+{3773\over 228\sqrt{6}}p^{(6)}\pa^4
\nonumber\\
&
+{3685\over 152\sqrt{6}}p^{(5)}\pa^5
+{11107\over
 456\sqrt{6}}p^{(4)}\pa^6
+{827\over 152}\sqrt{3\over2}p^{(3)}\pa^7
+{317\over 48\sqrt{6}}p''\pa^8
+{19871\over 19152\sqrt{6}}p'\pa^9
\nonumber\\
&
-{505\over 2394\sqrt{6}}p\pa^{10}
-{229\over 9455616\sqrt{6}}\pa^{28}
\Bigr) \psi_1
+\Bigl(
-{1\over4\sqrt{3}}p^{(5)}-{1\over2}\sqrt{3}p^{(4)}\pa
-{17\over 4\sqrt{3}}p^{(3)}\pa^2
\nonumber\\
& 
-{5\over2}\sqrt{3}p'' \pa^3
-{457\over 56\sqrt{3}}p'\pa^4-{29\over7\sqrt{3}}\pa^5
-{7\over 27648\sqrt{3}}\pa^{23}
\Bigr)\psi_7
+\Bigl(p'+4p \pa
\Bigr)\tilde{\psi}_{16}=0.
\label{eq:e7-3}
\end{align}
Eliminating $\tilde{\psi}_{16}$ and $\psi_7$, we obtain the differential equation including the pseudo-differential operator for the top component $\psi_1$ of $\Psi$.

For $E_8$, from the 248-dimensional adjoint representation whose structure constants can be 
evaluated  in \cite{gil,vav1},  the linear problem
(\ref{eq:linear problem}) can be formulated. 
Since it is rather complicated, we do not write down
the (pseudo-)ODE here. 
But for the completeness the generators in the adjoint representation of $E_8$ are presented in the appendix. 

From these pseudo-ODEs we expect that the Wronskians of the solutions satisfy the
E-type T-/Y-systems, which is left for future study.

\section{Conclusions and Discussions}
In this paper, we studied the relation between the quantum spectral curves of the Argyres-Douglas theories  and the integrable systems represented by the functional relations. 
For  a class of the AD theories without flavor symmetry, the quantum spectral curves become the ordinary differential equations  which  appear in the study of the ODE/IM correspondence.
We used the ODE/IM correspondence to obtain the T-system and Y-system satisfied by the Wronskians of the solutions. 
In particular we  have studied in detail for $(A_{1},A_n)$-type ODE with even $n$ and found that the corresponding integrable models are realized by the non-unitary minimal models as predicted by the 4d/2d correspondence of the AD theories. This was generalized to a class of $(A_m,A_n)$-type ODEs. Thus the ODE/IM correspondence provides a useful tool to study the AD theories explicitly from the viewpoint of the integrable systems. In particular, we have shown an explicit connection between the Y-functions  and the quantum periods of the $(A_1,A_n)$-type AD theories.
The quantum periods can be expanded analytically in terms of  the integrals of motion of the 2d CFT \cite{Bazhanov:1994ft,Bazhanov:1996aq}. 
 It is an interesting problem to compare the higher order corrections to the quantum periods in the WKB approximation and the one obtained from the TBA equations in order to see the connection between the AD theories and the 2d integrable system more precisely.
It is also interesting to generalize the present massless ODE/IM correspondence to the massive ODE/IM correspondence \cite{Lukyanov:2010rn}, which describes
the flow between two distinct AD theories via the integrable deformations.

For odd $n$, the $(A_1, A_n)$-type AD theories have flavor symmetries \cite{Wang:2015mra} and the related 2d CFT has the affine Lie algebra symmetry. 
It is interesting to study the corresponding 2d CFT from the viewpoint of the ODE/IM correspondence \cite{Feigin:2007mr}.

The quantum spectral curves for the $(A_m,A_n)$-type AD theories can be regarded as the differential equations obtained from the conformal limit of the linear problem associated with the modified $A_m^{(1)}$ affine Toda field equations. 
Based on this construction  we propose the quantum spectral curve for $({\mathfrak g},A_1)$-type AD theories for simply-laced Lie algebra ${\mathfrak g}$.
However the (pseudo-)ODE is rather complicated.
So it would be better to study the T-/Y-systems and the TBA equations from the Wronskian of the solutions of the linear system \cite{Lukyanov:2010rn,Ito:2016qzt}.

\subsection*{Acknowledgments}
We would like to thank  K.Maruyoshi, H.Nakajima, J. Suzuki, B. Vicedo, G.Yang, P. Zhao and Y. Zhou  for useful discussions. The work of KI is supported in part by Grant-in-Aid for Scientific Research 15K05043 and 16F16735 from Japan Society for the Promotion of Science (JSPS). 
The work of HS was supported by JSPS KAKENHI Grant Number 17J07135.

\appendix
\section{Matrix Representation for $E_r$-type Lie algebras}\label{sec:E_6-27-rep}
In this appendix we summarize the matrix representations for the $E_r$-type Lie algebras. 
In particular, we give the explicit matrix representation for the Chevalley generator $ E_{\alpha_i}$ ($i=0,\ldots,r$). 
Other generators $E_{-\alpha_i}$ and $\alpha_i^{\vee}\cdot H$ can be expressed 
as
\bee
E_{-\alpha_i}={}^tE_{\alpha_i},~~~\alpha_i^{\vee}\cdot H=[E_{\alpha_i},E_{-\alpha_i}].
\label{eq:other}
\ee
In the following $E_{i,j}$ denotes the matrix with components $\delta_{ia}\delta_{jb}$.

For $E_6$, 
the highest root is
$
\theta=\alpha_{1}+2\alpha_{2}+3\alpha_{3}+2\alpha_{4}+\alpha_{5}+2\alpha_{6}. 
$
In the 27-dimensional representation, the  generators $E_{\alpha_i}$ are given by
\begin{align}
 E_{\alpha_{1}}&=E_{1,2}+E_{12,15}+E_{14,17}+E_{16,19}+E_{18,21}+E_{20,22}, \nonumber\\
E_{\alpha_{2}}&=E_{2,3}+E_{10,12}+E_{11,14}+E_{13,16}+E_{21,23}+E_{22,24}, \nonumber\\
E_{\alpha_{3}}&=E_{3,4}+E_{8,10}+E_{9,11}+E_{16,18}+E_{19,21}+E_{24,25}, \nonumber\\
E_{\alpha_{4}}&=E_{4,5}+E_{6,8}+E_{11,13}+E_{14,16}+E_{17,19}+E_{25,26},
 \nonumber\\
E_{\alpha_{5}}&=E_{5,7}+E_{8,9}+E_{10,11}+E_{12,14}+E_{15,17}+E_{26,27}, \nonumber\\
E_{\alpha_{6}}&=E_{4,6}+E_{5,8}+E_{7,9}+E_{18,20}+E_{21,22}+E_{23,24}, 
\nonumber\\
E_{\alpha_0}&=
E_{20,1}+E_{22,1}+E_{24,3}+E_{25,4}+E_{26,5}+E_{27,7}.
\end{align}

For $E_7$,
the highest root is $\theta=2\alpha_1+3\alpha_2+4\alpha_3+3\alpha_4+2\alpha_5+\alpha_6+2\alpha_7$.
In the 56-dimensional representation, the generators $E_{\alpha_i}$ are given by
\begin{align}
E_{\alpha_1}&=E_{7,8}+E_{9,10}+E_{11,12}+E_{13,15}+E_{16,18}+E_{19,22}+E_{35,38}
\nonumber\\
&+E_{39,41}+E_{42,44}+E_{45,46}+E_{47,48}+E_{49,50},
\end{align}
\begin{align}
E_{\alpha_2}&=
E_{5,6} + E_{7,9} + E_{8,10} + E_{20,23} + E_{24,26} + E_{27,29} + E_{28,30} 
\nonumber\\
&+ E_{31,33} + E_{34,37} + E_{47,49} + E_{48,50} + E_{51,52},
\end{align}
\begin{align}
E_{\alpha_3}&=E_{5,7} + E_{6,9} + E_{12,14} + E_{15,17} + E_{18,21} +
 E_{22,25} 
\nonumber\\
&+ E_{32,35} + E_{36,39} + E_{40,42} + E_{43,45} + E_{48,51} + E_{50,52},
\end{align}
\begin{align}
E_{\alpha_4}&=E_{4,5} + E_{9,11} + E_{10,12} + E_{17,20} + E_{21,24} +
 E_{25,28} 
\nonumber\\
&+ E_{29,32} + E_{33,36} + E_{37,40} + E_{45,47} + E_{46,48} + E_{52,53},
\end{align}
\begin{align}
E_{\alpha_5}&=E_{3,4} + E_{11,13} + E_{12,15} + E_{14,17} + E_{24,27} +
 E_{26,29} 
\nonumber\\
&+ E_{28,31} + E_{30,33} + E_{40,43} + E_{42,45} + E_{44,46} + E_{53,54},
\end{align}
\begin{align}
E_{\alpha_6}&=E_{2,3} + E_{13,16} + E_{15,18} + E_{17,21} + E_{20,24} +
 E_{23,26} 
\nonumber\\
&+ E_{31,34} + E_{33,37} + E_{36,40} + E_{39,42} + E_{41,44} + E_{54,55},
\end{align}
\begin{align}
E_{\alpha_7}&=E_{1,2} + E_{16,19} + E_{18,22} + E_{21,25} + E_{24,28} +
 E_{26,30} 
\nonumber\\
&+ E_{27,31} + E_{29,33} + E_{32,36} + E_{35,39} + E_{38,41} + E_{55,56},
\end{align}
\begin{align}
E_{\alpha_0}&=E_{38,1}+E_{41,2}+E_{44,3}+E_{46,4}+E_{48,5}
+E_{50,6}\nonumber\\
&
+E_{51,7}+E_{52,9}+E_{53,11}+E_{54,13}+E_{55,16}+E_{56,19}.
\end{align} 

For the Lie algebra $E_8$ the highest root is given by
$\theta=2\alpha_1+3\alpha_2+4\alpha_3+5\alpha_4+6\alpha_5+4\alpha_6+2\alpha_7+3\alpha_8$.
In the 248-dimensional representation, the generators $E_{\alpha_i}$ are given by
\begin{align}
E_{\alpha_1}&=2 E_{1, 9} - E_{2, 9} - E_{10, 17} - E_{18, 24} - E_{25, 31} - 
 E_{32, 38} - E_{39, 45} - E_{40, 46} - E_{47, 52} - E_{48, 53}
\nonumber\\
& - 
 E_{54, 59} - E_{55, 60} - E_{61, 65} - E_{62, 66} - E_{67, 71} - 
 E_{68, 72} - E_{69, 73} - E_{74, 78} - E_{75, 79} - E_{80, 84}\nonumber\\
& - 
 E_{81, 85} - E_{86, 89} - E_{87, 90} - E_{91, 94} - E_{92, 95} - 
 E_{96, 99} - E_{100, 103} - E_{104, 107} - E_{108, 111}\nonumber\\
& - 
 E_{127, 128} - E_{129, 1} + E_{137, 130} + E_{144, 138} + 
 E_{151, 145} + E_{158, 152} + E_{165, 159} + E_{166, 160}\nonumber\\
& + 
 E_{172, 167} + E_{173, 168} + E_{179, 174} + E_{180, 175} + 
 E_{185, 181} + E_{186, 182} + E_{191, 187} + E_{192, 188}\nonumber\\
& + 
 E_{193, 189} + E_{198, 194} + E_{199, 195} + E_{204, 200} + 
 E_{205, 201} + E_{209, 206} + E_{210, 207} + E_{214, 211}\nonumber\\
& + 
 E_{215, 212} + E_{219, 216} + E_{223, 220} + E_{227, 224} + 
 E_{231, 228} + E_{248, 247},
\end{align}
\begin{align}
E_{\alpha_2}&=
-E_{1, 10} + 2 E_{2, 10} - E_{3, 10} + E_{9, 17} - E_{11, 18} - 
 E_{19, 25} - E_{26, 32} - E_{33, 39} - E_{34, 40} - E_{41, 47}\nonumber\\
& - 
 E_{42, 48} - E_{49, 54} - E_{50, 55} - E_{56, 61} - E_{57, 62} - 
 E_{63, 68} - E_{64, 69} - E_{70, 75} - E_{71, 77} - E_{76, 81}\nonumber\\
& - 
 E_{78, 83} - E_{82, 87} - E_{84, 88} - E_{89, 93} - E_{94, 97} - 
 E_{95, 98} - E_{99, 102} - E_{103, 106} - E_{107, 110} \nonumber\\
&- 
 E_{111, 114} - E_{126, 127} - E_{130, 2} - E_{137, 129} + 
 E_{138, 131} + E_{145, 139} + E_{152, 146} + E_{159, 153} \nonumber\\
&+ 
 E_{160, 154} + E_{167, 161} + E_{168, 162} + E_{174, 169} + 
 E_{175, 170} + E_{181, 176} + E_{182, 177} + E_{188, 183}\nonumber\\
& + 
 E_{189, 184} + E_{195, 190} + E_{197, 191} + E_{201, 196} + 
 E_{203, 198} + E_{207, 202} + E_{208, 204} + E_{213, 209}\nonumber\\
& + 
 E_{217, 214} + E_{218, 215} + E_{222, 219} + E_{226, 223} + 
 E_{230, 227} + E_{234, 231} + E_{247, 246},
\end{align}
\begin{align}
E_{\alpha_3}&=-E_{2, 11} + 2 E_{3, 11} - E_{4, 11} + E_{10, 18} - E_{12, 19} + 
 E_{17, 24} - E_{20, 26} - E_{27, 33} - E_{28, 34} - E_{35, 41}\nonumber\\
& - 
 E_{36, 42} - E_{43, 49} - E_{44, 50} - E_{51, 57} - E_{58, 64} - 
 E_{61, 67} - E_{65, 71} - E_{68, 74} - E_{72, 78} - E_{75, 80}\nonumber\\
& - 
 E_{79, 84} - E_{81, 86} - E_{85, 89} - E_{87, 92} - E_{90, 95} - 
 E_{97, 101} - E_{102, 105} - E_{106, 109} - E_{110, 113}\nonumber\\
& - 
 E_{114, 116} - E_{125, 126} - E_{131, 3} - E_{138, 130} + 
 E_{139, 132} - E_{144, 137} + E_{146, 140} + E_{153, 147}\nonumber\\
& + 
 E_{154, 148} + E_{161, 155} + E_{162, 156} + E_{169, 163} + 
 E_{170, 164} + E_{177, 171} + E_{184, 178} + E_{187, 181}\nonumber\\
& + 
 E_{191, 185} + E_{194, 188} + E_{198, 192} + E_{200, 195} + 
 E_{204, 199} + E_{206, 201} + E_{209, 205} + E_{212, 207}\nonumber\\
& + 
 E_{215, 210} + E_{221, 217} + E_{225, 222} + E_{229, 226} + 
 E_{233, 230} + E_{236, 234} + E_{246, 245},
\end{align}
\begin{align}
E_{\alpha_4}&=-E_{3, 12} + 2 E_{4, 12} - E_{5, 12} + E_{11, 19} - E_{13, 20} + 
 E_{18, 25} - E_{21, 27} - E_{22, 28} + E_{24, 31} - E_{29, 35}\nonumber\\
& - 
 E_{30, 36} - E_{37, 44} - E_{49, 56} - E_{54, 61} - E_{57, 63} - 
 E_{59, 65} - E_{62, 68} - E_{64, 70} - E_{66, 72} - E_{69, 75}\nonumber\\
& - 
 E_{73, 79} - E_{86, 91} - E_{89, 94} - E_{92, 96} - E_{93, 97} - 
 E_{95, 99} - E_{98, 102} - E_{109, 112} - E_{113, 115}\nonumber\\
& - 
 E_{116, 118} - E_{124, 125} - E_{132, 4} - E_{139, 131} + 
 E_{140, 133} - E_{145, 138} + E_{147, 141} + E_{148, 142}\nonumber\\
& - 
 E_{151, 144} + E_{155, 149} + E_{156, 150} + E_{164, 157} + 
 E_{176, 169} + E_{181, 174} + E_{183, 177} + E_{185, 179}\nonumber\\
& + 
 E_{188, 182} + E_{190, 184} + E_{192, 186} + E_{195, 189} + 
 E_{199, 193} + E_{211, 206} + E_{214, 209} + E_{216, 212}\nonumber\\
& + 
 E_{217, 213} + E_{219, 215} + E_{222, 218} + E_{232, 229} + 
 E_{235, 233} + E_{238, 236} + E_{245, 244},
\end{align}
\begin{align}
E_{\alpha_5}&=-E_{4, 13} + 2 E_{5, 13} - E_{6, 13} - E_{8, 13} + E_{12, 20} - 
 E_{14, 21} - E_{16, 22} + E_{19, 26} - E_{23, 29} + E_{25, 32}\nonumber\\
& + 
 E_{31, 38} - E_{36, 43} - E_{42, 49} - E_{44, 51} - E_{48, 54} - 
 E_{50, 57} - E_{53, 59} - E_{55, 62} - E_{60, 66} - E_{70, 76} \nonumber\\
&- 
 E_{75, 81} - E_{79, 85} - E_{80, 86} - E_{84, 89} - E_{88, 93} - 
 E_{96, 100} - E_{99, 103} - E_{102, 106} - E_{105, 109} \nonumber\\
&- 
 E_{115, 117} - E_{118, 119} - E_{123, 124} - E_{133, 5} - 
 E_{140, 132} + E_{141, 134} + E_{142, 136} - E_{146, 139}\nonumber\\
& + 
 E_{149, 143} - E_{152, 145} - E_{158, 151} + E_{163, 156} + 
 E_{169, 162} + E_{171, 164} + E_{174, 168} + E_{177, 170}\nonumber\\
& + 
 E_{179, 173} + E_{182, 175} + E_{186, 180} + E_{196, 190} + 
 E_{201, 195} + E_{205, 199} + E_{206, 200} + E_{209, 204}\nonumber\\
& + 
 E_{213, 208} + E_{220, 216} + E_{223, 219} + E_{226, 222} + 
 E_{229, 225} + E_{237, 235} + E_{239, 238} + E_{244, 243},
\end{align}

\begin{align}
E_{\alpha_6}&=-E_{5, 14} + 2 E_{6, 14} - E_{7, 14} + E_{13, 21} - E_{15, 23} + 
 E_{20, 27} - E_{22, 30} + E_{26, 33} - E_{28, 36} + E_{32, 39}\nonumber\\
& - 
 E_{34, 42} + E_{38, 45} - E_{40, 48} - E_{46, 53} - E_{51, 58} - 
 E_{57, 64} - E_{62, 69} - E_{63, 70} - E_{66, 73} - E_{68, 75}\nonumber\\
& - 
 E_{72, 79} - E_{74, 80} - E_{78, 84} - E_{83, 88} - E_{100, 104} - 
 E_{103, 107} - E_{106, 110} - E_{109, 113} - E_{112, 115} \nonumber\\
&- 
 E_{119, 121} - E_{122, 123} - E_{134, 6} - E_{141, 133} + 
 E_{143, 135} - E_{147, 140} + E_{150, 142} - E_{153, 146}\nonumber\\
& + 
 E_{156, 148} - E_{159, 152} + E_{162, 154} - E_{165, 158} + 
 E_{168, 160} + E_{173, 166} + E_{178, 171} + E_{184, 177}\nonumber\\
& + 
 E_{189, 182} + E_{190, 183} + E_{193, 186} + E_{195, 188} + 
 E_{199, 192} + E_{200, 194} + E_{204, 198} + E_{208, 203}\nonumber\\
& + 
 E_{224, 220} + E_{227, 223} + E_{230, 226} + E_{233, 229} + 
 E_{235, 232} + E_{241, 239} + E_{243, 242},
\end{align}

\begin{align}
E_{\alpha_7}&=-E_{6, 15} + 2 E_{7, 15} + E_{14, 23} + E_{21, 29} + E_{27, 35} - 
 E_{30, 37} + E_{33, 41} - E_{36, 44} + E_{39, 47} - E_{42, 50}\nonumber\\
& - 
 E_{43, 51} + E_{45, 52} - E_{48, 55} - E_{49, 57} - E_{53, 60} - 
 E_{54, 62} - E_{56, 63} - E_{59, 66} - E_{61, 68} - E_{65, 72}\nonumber\\
& - 
 E_{67, 74} - E_{71, 78} - E_{77, 83} - E_{104, 108} - E_{107, 111} - 
 E_{110, 114} - E_{113, 116} - E_{115, 118} - E_{117, 119}\nonumber\\
& - 
 E_{120, 122} - E_{135, 7} - E_{143, 134} - E_{149, 141} - 
 E_{155, 147} + E_{157, 150} - E_{161, 153} + E_{164, 156}\nonumber\\
& - 
 E_{167, 159} + E_{170, 162} + E_{171, 163} - E_{172, 165} + 
 E_{175, 168} + E_{177, 169} + E_{180, 173} + E_{182, 174}\nonumber\\
& + 
 E_{183, 176} + E_{186, 179} + E_{188, 181} + E_{192, 185} + 
 E_{194, 187} + E_{198, 191} + E_{203, 197} + E_{228, 224}\nonumber\\
& + 
 E_{231, 227} + E_{234, 230} + E_{236, 233} + E_{238, 235} + 
 E_{239, 237} + E_{242, 240},
\end{align}
\begin{align}
E_{\alpha_8}&=-E_{5, 16} + 2 E_{8, 16} + E_{13, 22} + E_{20, 28} - E_{21, 30} + 
 E_{26, 34} - E_{27, 36} + E_{29, 37} + E_{32, 40} - E_{33, 42}\nonumber\\
& + 
 E_{35, 44} + E_{38, 46} - E_{39, 48} + E_{41, 50} - E_{45, 53} + 
 E_{47, 55} + E_{52, 60} - E_{76, 82} - E_{81, 87} - E_{85, 90}\nonumber\\
& - 
 E_{86, 92} - E_{89, 95} - E_{91, 96} - E_{93, 98} - E_{94, 99} - 
 E_{97, 102} - E_{101, 105} - E_{117, 120} - E_{119, 122}\nonumber\\
& - 
 E_{121, 123} - E_{136, 8} - E_{142, 133} - E_{148, 140} + 
 E_{150, 141} - E_{154, 146} + E_{156, 147} - E_{157, 149}\nonumber\\
& - 
 E_{160, 152} + E_{162, 153} - E_{164, 155} - E_{166, 158} + 
 E_{168, 159} - E_{170, 161} + E_{173, 165} - E_{175, 167}\nonumber\\
& - 
 E_{180, 172} + E_{202, 196} + E_{207, 201} + E_{210, 205} + 
 E_{212, 206} + E_{215, 209} + E_{216, 211} + E_{218, 213}\nonumber\\
& + 
 E_{219, 214} + E_{222, 217} + E_{225, 221} + E_{240, 237} + 
 E_{242, 239} + E_{243, 241},
\end{align}
\begin{align}
E_{\alpha_0}&=-E_{1, 248} - E_{9, 247} - E_{17, 246} - E_{24, 245} - E_{31, 244} - 
 E_{38, 243} - E_{45, 242} - E_{46, 241} - E_{52, 240} \nonumber\\
&+ E_{53, 239} -
  E_{59, 238} - E_{60, 237} + E_{65, 236} + E_{66, 235} - 
 E_{71, 234} - E_{72, 233} - E_{73, 232} + E_{77, 231} \nonumber\\
&+ E_{78, 230} +
  E_{79, 229} - E_{83, 227} - E_{84, 226} - E_{85, 225} + 
 E_{88, 223} + E_{89, 222} + E_{90, 221} - E_{93, 219}\nonumber\\
& - E_{94, 218} -
  E_{95, 217} + E_{97, 215} + E_{98, 214} + E_{99, 213} - 
 E_{101, 210} - E_{102, 209} - E_{103, 208}\nonumber\\
&+ E_{105, 205} + 
 E_{106, 204} + E_{107, 203} - E_{109, 199} - E_{110, 198} - 
 E_{111, 197} + E_{112, 193} + E_{113, 192} \nonumber\\
& + E_{114, 191}- 
 E_{115, 186} - E_{116, 185} + E_{117, 180} + E_{118, 179} - 
 E_{119, 173} + E_{120, 172} + E_{121, 166} \nonumber\\
&+ E_{122, 165} + 
 E_{123, 158} + E_{124, 151} + E_{125, 144} + E_{126, 137} + 
 E_{127, 129} + 2 E_{128, 1} + 3 E_{128, 2} \nonumber\\
&+ 4 E_{128, 3} + 
 5 E_{128, 4} + 6 E_{128, 5} + 4 E_{128, 6} + 2 E_{128, 7} + 
 3 E_{128, 8}.
\end{align}


\begin{thebibliography}{99}

\bibitem{Argyres:1995jj} 
  P.~C.~Argyres and M.~R.~Douglas,
``New phenomena in SU(3) supersymmetric gauge theory,''
  Nucl.\ Phys.\ B {\bf 448}, 93 (1995)
  [hep-th/9505062].


\bibitem{Argyres:1995xn} 
  P.~C.~Argyres, M.~R.~Plesser, N.~Seiberg and E.~Witten,
``New N=2 superconformal field theories in four-dimensions,''
  Nucl.\ Phys.\ B {\bf 461}, 71 (1996)
  [hep-th/9511154].


\bibitem{Eguchi:1996vu} 
  T.~Eguchi, K.~Hori, K.~Ito and S.~K.~Yang,
``Study of N=2 superconformal field theories in four-dimensions,''
  Nucl.\ Phys.\ B {\bf 471}, 430 (1996)
  [hep-th/9603002].


\bibitem{Wang:2015mra} 
  Y.~Wang and D.~Xie,
``Classification of Argyres-Douglas theories from M5 branes,''
  Phys.\ Rev.\ D {\bf 94}, no. 6, 065012 (2016)
  [arXiv:1509.00847 [hep-th]].


\bibitem{Xie:2012hs} 
  D.~Xie,
``General Argyres-Douglas Theory,''
  JHEP {\bf 1301}, 100 (2013)
  [arXiv:1204.2270 [hep-th]].



\bibitem{Beem:2013sza} 
  C.~Beem, M.~Lemos, P.~Liendo, W.~Peelaers, L.~Rastelli and B.~C.~van Rees,
``Infinite Chiral Symmetry in Four Dimensions,''
  Commun.\ Math.\ Phys.\  {\bf 336}, no. 3, 1359 (2015)
  [arXiv:1312.5344 [hep-th]].

\bibitem{Cordova:2015nma} 
  C.~Cordova and S.~H.~Shao,
``Schur Indices, BPS Particles, and Argyres-Douglas Theories,''
  JHEP {\bf 1601}, 040 (2016)
  [arXiv:1506.00265 [hep-th]].
  
\bibitem{Buican:2015ina} 
  M.~Buican and T.~Nishinaka,
``On the superconformal index of Argyres-Douglas theories,''
  J.\ Phys.\ A {\bf 49}, no. 1, 015401 (2016)
  [arXiv:1505.05884 [hep-th]].


\bibitem{Xie:2016evu} 
  D.~Xie, W.~Yan and S.~T.~Yau,
``Chiral algebra of Argyres-Douglas theory from M5 brane,''
  arXiv:1604.02155 [hep-th].


\bibitem{Fredrickson:2017yka} 
  L.~Fredrickson, D.~Pei, W.~Yan and K.~Ye,
``Argyres-Douglas Theories, Chiral Algebras and Wild Hitchin Characters,''
  arXiv:1701.08782 [hep-th].


\bibitem{Nekrasov:2009rc} 
  N.~A.~Nekrasov and S.~L.~Shatashvili,
``Quantization of Integrable Systems and Four Dimensional Gauge Theories,''
  arXiv:0908.4052 [hep-th].


\bibitem{Mironov:2009uv} 
  A.~Mironov and A.~Morozov,
``Nekrasov Functions and Exact Bohr-Zommerfeld Integrals,''
  JHEP {\bf 1004}, 040 (2010)
  [arXiv:0910.5670 [hep-th]].


\bibitem{Gaiotto:2014bza}
  D.~Gaiotto,
  ``Opers and TBA,''
  arXiv:1403.6137 [hep-th].


\bibitem{Cecotti:2014zga} 
  S.~Cecotti and M.~Del Zotto,
``$Y$ systems, $Q$ systems, and 4D $\mathcal{N}=2$ supersymmetric QFT,''
  J.\ Phys.\ A {\bf 47}, no. 47, 474001 (2014)
  [arXiv:1403.7613 [hep-th]].

\bibitem{Gaiotto:2008cd} 
  D.~Gaiotto, G.~W.~Moore and A.~Neitzke,
``Four-dimensional wall-crossing via three-dimensional field theory,''
  Commun.\ Math.\ Phys.\  {\bf 299}, 163 (2010)
  [arXiv:0807.4723 [hep-th]].
  

\bibitem{Dorey:1998pt} 
  P.~Dorey and R.~Tateo,
``Anharmonic oscillators, the thermodynamic Bethe ansatz, and nonlinear integral equations,''
  J.\ Phys.\ A {\bf 32}, L419 (1999)
  [hep-th/9812211].


\bibitem{Bazhanov:1998wj} 
  V.~V.~Bazhanov, S.~L.~Lukyanov and A.~B.~Zamolodchikov,
``Spectral determinants for Schrodinger equation and Q operators of conformal field theory,''
  J.\ Statist.\ Phys.\  {\bf 102}, 567 (2001)
  [hep-th/9812247].


\bibitem{Dorey:2007zx} 
  P.~Dorey, C.~Dunning and R.~Tateo,
``The ODE/IM Correspondence,''
  J.\ Phys.\ A {\bf 40}, R205 (2007)
  [hep-th/0703066].


\bibitem{Dorey:2006an} 
  P.~Dorey, C.~Dunning, D.~Masoero, J.~Suzuki and R.~Tateo,
``Pseudo-differential equations, and the Bethe ansatz for the classical Lie algebras,''
  Nucl.\ Phys.\ B {\bf 772}, 249 (2007)
  [hep-th/0612298].


\bibitem{Lukyanov:2010rn} 
  S.~L.~Lukyanov and A.~B.~Zamolodchikov,
``Quantum Sine(h)-Gordon Model and Classical Integrable Equations,''
  JHEP {\bf 1007}, 008 (2010)
  [arXiv:1003.5333 [math-ph]].


\bibitem{Dorey:2012bx} 
  P.~Dorey, S.~Faldella, S.~Negro and R.~Tateo,
``The Bethe Ansatz and the Tzitzeica-Bullough-Dodd equation,''
  Phil.\ Trans.\ Roy.\ Soc.\ Lond.\ A {\bf 371}, 20120052 (2013)
  [arXiv:1209.5517 [math-ph]].


\bibitem{Ito:2013aea} 
  K.~Ito and C.~Locke,
``ODE/IM correspondence and modified affine Toda field equations,''
  Nucl.\ Phys.\ B {\bf 885}, 600 (2014)
  [arXiv:1312.6759 [hep-th]].


\bibitem{Adamopoulou:2014fca} 
  P.~Adamopoulou and C.~Dunning,
``Bethe Ansatz equations for the classical $A_n^{(1)}$ affine Toda field theories,''
  J.\ Phys.\ A {\bf 47}, 205205 (2014)
  [arXiv:1401.1187 [math-ph]].


\bibitem{Ito:2015nla} 
  K.~Ito and C.~Locke,
``ODE/IM correspondence and Bethe ansatz for affine Toda field equations,''
  Nucl.\ Phys.\ B {\bf 896}, 763 (2015)
  [arXiv:1502.00906 [hep-th]].


\bibitem{Masoero:2015lga} 
  D.~Masoero, A.~Raimondo and D.~Valeri,
``Bethe Ansatz and the Spectral Theory of Affine Lie Algebra-Valued Connections I. The simply-laced Case,''
  Commun.\ Math.\ Phys.\  {\bf 344}, no. 3, 719 (2016)
  [arXiv:1501.07421 [math-ph]].


\bibitem{Masoero:2015rcz} 
  D.~Masoero, A.~Raimondo and D.~Valeri,
``Bethe Ansatz and the Spectral Theory of Affine Lie algebra-Valued Connections II: The Non Simply-Laced Case,''
  Commun.\ Math.\ Phys.\  {\bf 349}, no. 3, 1063 (2017)
  [arXiv:1511.00895 [math-ph]].


\bibitem{Sun:2012xw} 
  J.~Sun,
``Polynomial relations for $q$-characters via the ODE/IM correspondence,''
  SIGMA {\bf 8}, 028 (2012)
  [arXiv:1201.1614 [math.QA]].


\bibitem{Zamolodchikov:1991et} 
  A.~B.~Zamolodchikov,
``On the thermodynamic Bethe ansatz equations for reflectionless ADE scattering theories,''
  Phys.\ Lett.\ B {\bf 253}, 391 (1991).

\bibitem{Klemm:1994qs} 
  A.~Klemm, W.~Lerche, S.~Yankielowicz and S.~Theisen,
  Phys.\ Lett.\ B {\bf 344}, 169 (1995)
  doi:10.1016/0370-2693(94)01516-F
  [hep-th/9411048].

\bibitem{Martinec:1995by} 
  E.~J.~Martinec and N.~P.~Warner,
``Integrable systems and supersymmetric gauge theory,''
  Nucl.\ Phys.\ B {\bf 459}, 97 (1996)
  [hep-th/9509161].


\bibitem{Ito:1999cc} 
  K.~Ito,
``A-D-E singularity and the Seiberg-Witten theory,''
  Prog.\ Theor.\ Phys.\ Suppl.\  {\bf 135}, 94 (1999)
  [hep-th/9906023].


\bibitem{Dorey:2000ma} 
  P.~Dorey, C.~Dunning and R.~Tateo,
``Differential equations for general SU(n) Bethe ansatz systems,''
  J.\ Phys.\ A {\bf 33}, 8427 (2000)
  [hep-th/0008039].

\bibitem{Sibuya}
Y.\ Sibuya,
 Global Theory of a second-order linear ordinary differential operator with polynomial coefficient
(Amsterdam: North-Holland 1975)

\bibitem{Bazhanov:1994ft} 
  V.~V.~Bazhanov, S.~L.~Lukyanov and A.~B.~Zamolodchikov,
``Integrable structure of conformal field theory, quantum KdV theory and thermodynamic Bethe ansatz,''
  Commun.\ Math.\ Phys.\  {\bf 177}, 381 (1996)
  [hep-th/9412229].


\bibitem{Mathieu:1990dy} 
  P.~Mathieu and M.~A.~Walton,
``Fractional level Kac-Moody algebras and nonunitarity coset conformal theories,''
  Prog.\ Theor.\ Phys.\ Suppl.\  {\bf 102}, 229 (1990).


\bibitem{Bazhanov:1996dr} 
  V.~V.~Bazhanov, S.~L.~Lukyanov and A.~B.~Zamolodchikov,
``Integrable structure of conformal field theory. 2. Q operator and DDV equation,''
  Commun.\ Math.\ Phys.\  {\bf 190}, 247 (1997)
  [hep-th/9604044].


\bibitem{Alday:2010vh} 
  L.~F.~Alday, J.~Maldacena, A.~Sever and P.~Vieira,
``Y-system for Scattering Amplitudes,''
  J.\ Phys.\ A {\bf 43}, 485401 (2010)
  [arXiv:1002.2459 [hep-th]].




\bibitem{Bazhanov:1996aq} 
  V.~V.~Bazhanov, S.~L.~Lukyanov and A.~B.~Zamolodchikov,
``Integrable quantum field theories in finite volume: Excited state energies,''
  Nucl.\ Phys.\ B {\bf 489}, 487 (1997)
  [hep-th/9607099].


\bibitem{Ravanini:1992fi} 
  F.~Ravanini, R.~Tateo and A.~Valleriani,
``Dynkin TBAs,''
  Int.\ J.\ Mod.\ Phys.\ A {\bf 8}, 1707 (1993)
  [hep-th/9207040].


\bibitem{Dorey:2007ti} 
  P.~Dorey, C.~Dunning, F.~Gliozzi and R.~Tateo,
``On the ODE/IM correspondence for minimal models,''
  J.\ Phys.\ A {\bf 41}, 132001 (2008)
  [arXiv:0712.2010 [hep-th]].


\bibitem{Dunning:2002cu} 
  C.~Dunning,
``Massless flows between minimal W models,''
  Phys.\ Lett.\ B {\bf 537}, 297 (2002)
  [hep-th/0204090].


\bibitem{gil}
P.B.Gilkey and G.M.Seitz,``Some representations of exceptional Lie algebras," Geometriae Dedicata {\bf 25} (1988) 407.

\bibitem{vav1} 
N.A. Vavilov, ``Do It Yourself: the Structure Constants for Lie Algebras of Types $E_l$," J.Math. Sci. {\bf 120} (2004) 1513.


\bibitem{Feigin:2007mr} 
  B.~Feigin and E.~Frenkel,
``Quantization of soliton systems and Langlands duality,''
  arXiv:0705.2486 [math.QA].


\bibitem{Creutzig:2017qyf} 
  T.~Creutzig,
``W-algebras for Argyres-Douglas theories,''
  arXiv:1701.05926 [hep-th].

\bibitem{Ito:2016qzt} 
  K.~Ito and H.~Shu,
  ``ODE/IM correspondence for modified $B_2^{(1)}$ affine Toda field equation,''
  Nucl.\ Phys.\ B {\bf 916}, 414 (2017)
  [arXiv:1605.04668 [hep-th]].
 
\end{thebibliography}
\end{document}